\documentclass[manuscript]{aastex}

\shorttitle{Energy release in the lower solar atmosphere of a solar flare}
\shortauthors{Sharykin and Kosovichev}

\begin{document}

\title{OBSERVATIONAL INVESTIGATION OF ENERGY RELEASE IN THE LOWER SOLAR ATMOSPHERE OF A SOLAR FLARE}

\author{I.N. Sharykin\altaffilmark{1,2}, V.M.~Sadykov\altaffilmark{3},
A.G. Kosovichev\altaffilmark{3,4}, S.~Vargas-Dominguez\altaffilmark{5},
I.V.~Zimovets\altaffilmark{1}}

\affil{Space Research Institute of RAS, Moscow 117997, Russia}

\altaffiltext{1}{Space Research Institute (IKI) of the Russian Academy of Science}
\altaffiltext{2}{Institute of Solar-Terrestrial Research (ISTP) of the Russian Academy of Science, Siberian Branch}
\altaffiltext{3}{New Jersey Institute of Technology}
\altaffiltext{4}{NASA Ames Research Center}
\altaffiltext{5}{Universidad Nacional de Colombia}


\begin{abstract}

We study flare processes in the lower solar atmosphere using observational data for a M1-class flare of June 12, 2014, obtained by New Solar Telescope (NST/BBSO) and Helioseismic Magnetic Imager (HMI/SDO). The main goal is to understand triggers and manifestations of the flare energy release in the lower layers of the solar atmosphere (the photosphere and chromosphere) using high-resolution optical observations and magnetic field measurements. We analyze optical images, HMI Dopplergrams and vector magnetograms, and use Non-Linear Force-Free Field (NLFFF) extrapolations for reconstruction of the magnetic topology. The NLFFF modelling reveals interaction of oppositely directed magnetic flux-tubes in the PIL. These two interacting magnetic flux tubes are observed as a compact sheared arcade along the PIL in the high-resolution broad-band continuum images from NST. In the vicinity of the PIL, the NST H$\rm_{\alpha}$observations reveal formation of a thin three-ribbon structure corresponding to the small-scale photospheric magnetic arcade. Presented observational results evidence in favor of location of the primary energy release site in the dense chromosphere where plasma is partially ionized in the region of strong electric currents concentrated near the polarity inversion line. Magnetic reconnection may be triggered by two interacting magnetic flux tubes with forming current sheet elongated along the PIL.

\end{abstract}
\keywords{Sun: flares; Sun: photosphere; Sun: chromosphere; Sun: corona; Sun: magnetic fields}

\section{INTRODUCTION}

Magnetic reconnection is believed to be the main mechanism of solar flare energy release. Various scenarios of magnetic reconnection discussed in many papers including flare models by \cite{Somov1974,Syrovatskii1976,Amari1999,Somov2000,Priest2000,Priest2002}. Most of these models assume that the reconnection process occurs in the low-density high-temperature corona. However, there are a lot of variants of magnetic field topologies where magnetic reconnection can be triggered. The most popular standard flare model \citep{Hirayama1974,Magara1996,Tsuneta1997} assumes formation of a quasi-vertical current sheet in the solar corona beneath of an upward moving plasmoid. In the framework of this model the primary energy release (magnetic reconnection in the current sheet) and electron acceleration also take place in the low-density corona. It is likely that a large fraction of eruptive solar flares follows this scenario \citep[e.g.][]{Liu2008,Krucker2008,Fletcher2011}. Obviously, non-eruptive events cannot be interpreted in the frame of the standard model. Thanks to the development of multiwavelength observations of the Sun, nowdays we are getting a lot of information about the real complexity of the flare phenomenon and inapplicability of the standard model. It has been argued that the magnetic reconnection process can be triggered not only in the low-density corona but also in deeper partially ionized layers of the solar atmosphere \citep[e.g.][]{Georgoulis2002, Chae2003}. Recently developed models of chromospheric magnetic reconnection \citep[e.g.][]{Leake2012, Leake2013, Leake2014, Ni2015} have direct applications to the physics of such observed phenomena at different scales like chromospheric jets, Ellerman bombs, and, perhaps, solar flares. However, it is unclear if the chromospheric reconnection may occur on the larger scale of solar flares. In this respect, investigation of the flare energy release in the lower solar atmosphere is particularly interesting.

It has been known since observations of \cite{Severnyi1958} that the solar flares appear first in the vicinity of the polarity inversion line (PIL) of the line-of-sight magnetic field, and that the flare emission spreads outside the PIL as the flare develops. Also, he found that a considerable gradient of the magnetic field across the PIL is required for the flare appearance. Due to the availability of various multispectral observations of the solar flares, today we are getting new knowledge about the flare processes near the PIL. The statistical analysis of SOHO/MDI line-of-sight magnetograms, presented in the work of \cite{Welsch2008} shows that the PIL with strong gradient of the magnetic field is associated with the flux emergence. The most recent statistical study made by \cite{Schrijver2016} illustrates that strong-field, high-gradient polarity inversion lines (SHILs) created during emergence of magnetic flux into active region are associated with X-class flares. \cite{Wang2015} demonstrated changing horizontal component of the magnetic field near the PIL, which could be connected with magnetic reconnection in the twisted magnetic flux tube. A detailed 3D modelling of the magnetic field presented in the work of \cite{Inoue2016} reveals conditions in the vicinity of the PIL favourable for magnetic reconnection and triggering solar flares. High resolution observations of the formation of H$\rm_{\alpha}$ ribbons surrounding the twisted magnetic flux rope near the PIL are shown in the work of \cite{Wang2015}. In this work authors show evidence that the flux rope elongated along the PIL in the low solar atmosphere becomes unstable following the enhancement of its twists. Thus, it is especially interesting to study flare processes in the vicinity of PIL in more details.

In this paper we investigate the M1.0 non-eruptive solar flare occurred on 12 June, 2014, started at around 21:00~UT (peak at 21:12~UT and ended at $\approx$21:30~UT) in active region NOAA 12087, located approximately 50 degrees South-East from the disk center. This event was selected for analysis due to the availability of various unique observations of the solar atmosphere made by Interface Region Imaging Spectrograph, IRIS \citep{DePontieu2014}, New Solar Telescope, NST \citep{Goode2012} in the Big Bear Solar Observatory (BBSO), Helioseismic Magnetic Imager, HMI \citep{Scherrer2012} and Atmospheric Imaging Assembly, AIA \citep{Lemen2012} instruments onboard Solar Dynamics Observatory, SDO \citep{Pesnell2012}. This flare showed an intriguing activity in the vicinity of the PIL, which was discussed in a couple of previous works \citep{Sadykov2015,Kumar2015}. IRIS and RHESSI observations of the selected flare were previously discussed in the paper \cite{Sadykov2015}. This work presents the detailed analysis of Doppler shift maps reconstructed for different lines, which are compared with the RHESSI X-ray images. It was concluded that chromospheric evaporation could be triggered not only by accelerated particles but by heat conduction flows from hot plasma. IRIS and AIA/SDO images also reveal energy release in the form of dynamic jet, which erupts from the deep layers of the solar atmosphere in the vicinity of the magnetic field polarity inversion line (PIL). In the work \citep{Kumar2015} authors claim that they found direct evidence of magnetic reconnection in the vicinity of the PIL between two small opposite polarity sunspots. Using NST H$_{\alpha}$ images they concluded that interaction between two J-shape loops lead to the formation of the twisted flux-rope elongated along the PIL. They also detected plasma inflows near the PIL that was recognised as a signature of the magnetic reconnection. The focus of our paper is on a detailed analysis of physical processes in the vicinity of the PIL in the low atmosphere during the solar flare using high resolution observations.

We present analysis of the H$\rm_{\alpha}$ line wings filtergrams and broadband TiO images (7057~\AA), which correspondingly show processes in the lower chromosphere and the photosphere. We will also use nonlinear force-free magnetic field (NLFFF) modelling of the magnetic field to explain some topological peculiarities and energy release of the flare. Magnetic reconnection or triggering of a solar flare in the lower solar atmosphere can be connected with plasma flows. To detect these and compare with other data we use the HMI Dopplergrams, which give us information about the speed of plasma flows along the line of sight.

The paper is divided into four sections. Section 2 describes the observational data. Section 3 is devoted to the analysis of NST images, electric currents and magnetic field topology (using the HMI data). Discussion and conclusions are presented in Section 4.

\section{OBSERVATIONAL DATA}


To investigate the fine spatial structure of the flare energy release site we use $H_{\alpha}$ data from Visible Imaging Spectrometer (VIS) and TiO filter (7057~\AA) images from Broadband Filter Imager (BFI) obtained at the 1.6 m New Solar Telescope \citep{Goode2012,Varsik2014} and available from the BBSO web site (http://bbso.njit.edu/). The new system of adaptive optics, large aperture and speckle-interferometry postprocessing \citep{Woeger2008} make it possible to obtain images of the Sun with high spatial resolution up to the diffraction limit. The pixel size of the VIS images is about 0.029$^{\prime\prime}$, which is approximately 3 times smaller than the telescope diffraction limit $\lambda_{H\alpha}/D\approx 0.084^{\prime\prime}$. The pixel size of the BFI TiO images is about 0.034$^{\prime\prime}$ with the diffraction limit $\approx 0.09^{\prime\prime}$. The time cadence between two subsequent $\rm H_{\alpha}$ line scans in 5 wavelength bands (6563 $\rm\AA$ $\pm$1.0, $\pm$0.8, $\pm$0.4 and the line core) is $\approx$15 seconds.

We study the magnetic field structure using observations of Helioseismic Magnetic Imager \citep{Scherrer2012}. The HMI makes spectropolarometric observations of the magneto-sensitive Fe~I line (6173\AA) by two different cameras in six wavelength channels covering the whole line profile. Calculations of the optical continuum intensity, line-of-sight magnetic field and Doppler velocity with the time cadence 45 s and the spatial resolution of $\sim 1^{\prime\prime}$ are based on the right and left circular polarization measurements. To obtain the full magnetic field vector with the time cadence 720 s HMI also make measurements of the linear polarization profile. The spatial resolution of vector magnetograms is the same as in the case of the LOS magnetograms.

The 3D structure of the magnetic field in the flare region was reconstructed using the NLFFF optimization code \citep{Wheatland2000} available from the package SolarSoft. For the boundary condition we used the disambiguated HMI vector magnetograms \citep{Centeno2014}.

\section{RESULTS}

\subsection{NST OBSERVATIONS}

\begin{figure}[pH]
\centering
\includegraphics[width=0.75\linewidth]{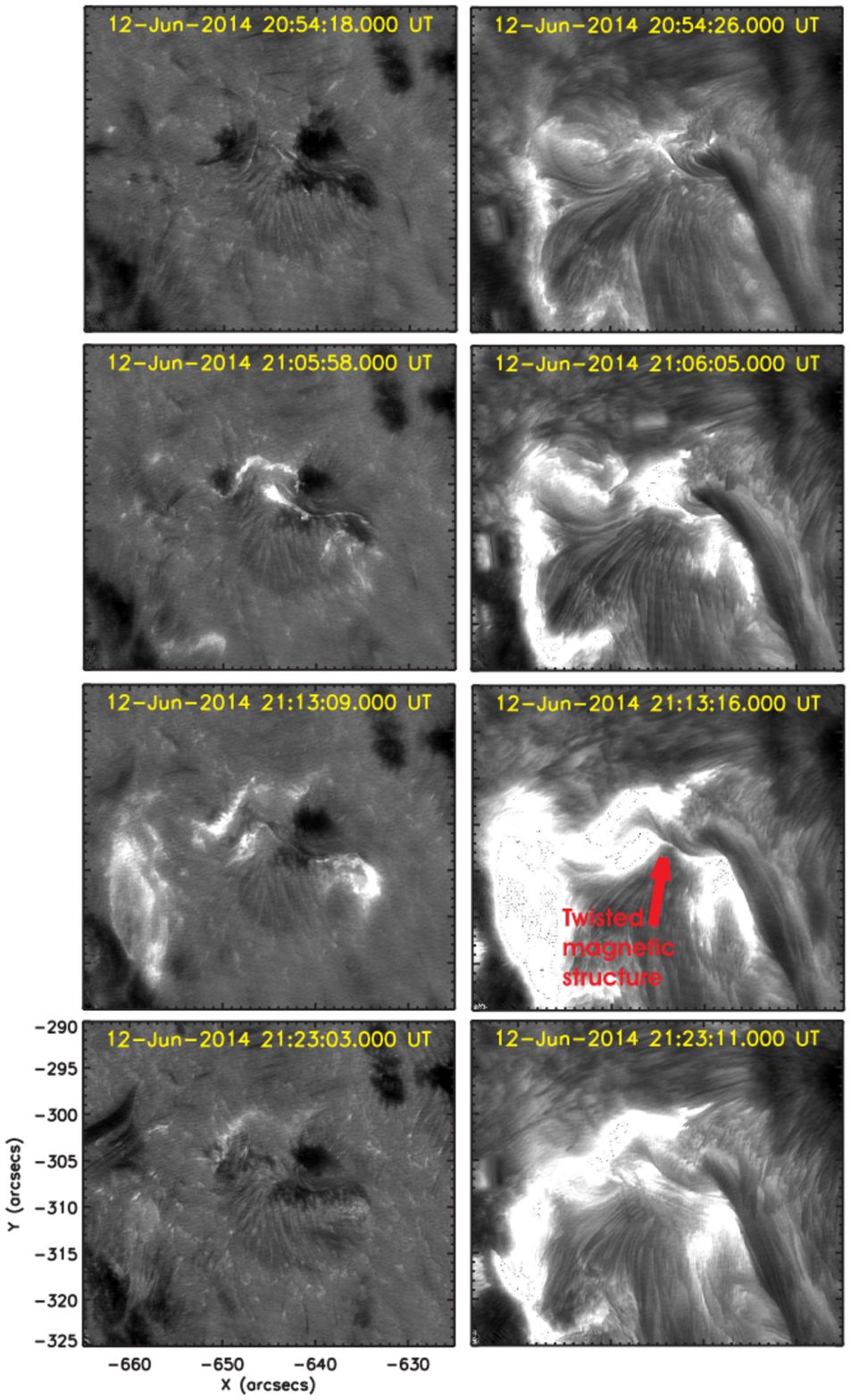}
\caption{Figure shows VIS/NST H$\rm_{\alpha}$ filtergrams in the red wing ($+0.8$~\AA, left panel) and in the line core (right panel) of a line. We marked the twisted magnetic structure seen in the H$\rm_{\alpha}$ line center filtergram by red arrow.}
\label{VIS_large}
\end{figure}

\begin{figure}[h!]
\centering
\includegraphics[width=1.0\linewidth]{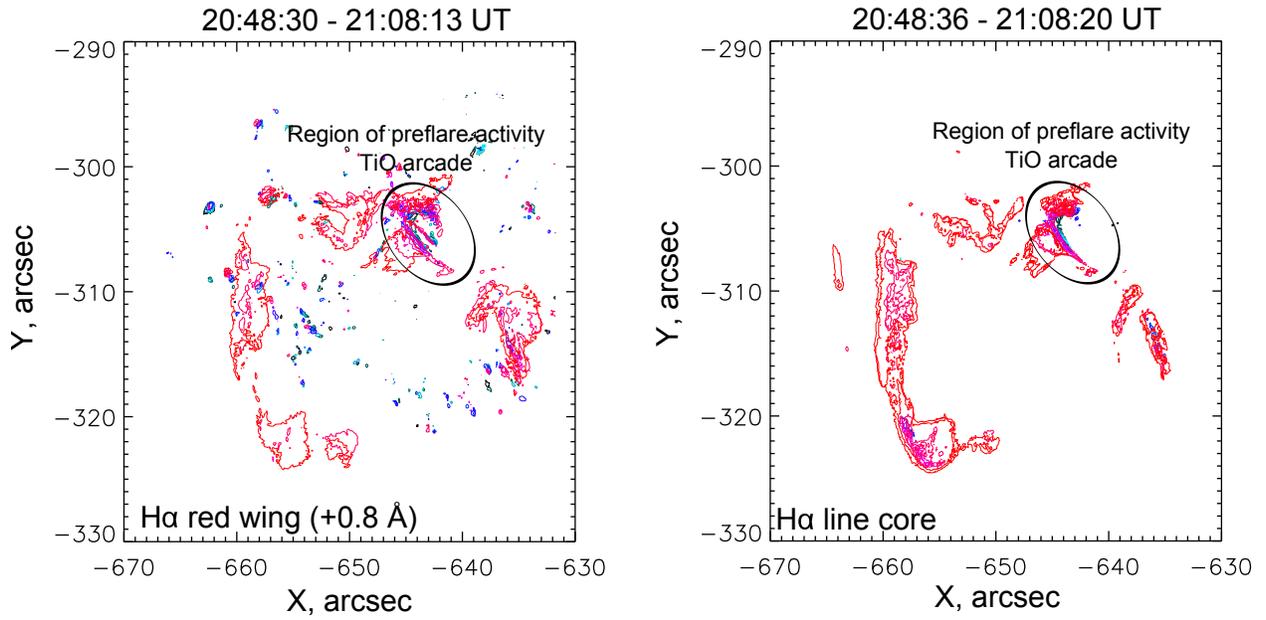}
\caption{Structure of the H$\rm_{\alpha}$ emission sources (marked by contours) in the whole flare region. The H$\rm_{\alpha}$ red wing (level is 13000 DNs) and line core (level is 26000 DNs) emission sources are shown in the left and right panels respectively. Blue lines correspond to the preflare time; red lines show the H$\rm_{\alpha}$ emission source during the flare time (the time range is shown in the upper part of the figures).}
\label{rib_evo}
\end{figure}

In Fig.\,\ref{VIS_large} we show a sequence of H$\rm_{\alpha}$~filtergrams (the line core and red wing~6563$+0.8\rm\AA$) for a large field of view (FOV). The $H_{\alpha}$ line wing filtergram corresponds to deep layers of the chromosphere, as the emission in this band experiences less absorption than in the H$\rm_{\alpha}$ line core. The line-core center filtergrams reveal the complexity of the upper chromospheric flare region. One can note a twisted structure (indicated by red arrow) passing through the image. The H$\rm_{\alpha}$ line core filtergrams show ribbon-like emission sources during the whole event, while emission in the line wings was observed only during a short time interval. The most interesting observation is that the initial emission is developed in a small region surrounding a twisted magnetic structure located in the vicinity of the polarity inversion line. According to the work of \cite{Sadykov2015} and \cite{Kumar2015} this place corresponds to a jet observed by IRIS, and probably corresponds to the place of magnetic reconnection. To show the flare build-up phase and locate the preflare energy release site in the low solar atmosphere we present H$\rm_{\alpha}$ contour images before the flare (blue tones) and during the flare (red tones) in Fig.\,\ref{rib_evo}. The most significant preflare activity was observed in the region marked by ellipse. Our further analysis will be mostly focused on this particular region.

\begin{figure}[pH]
\centering
\includegraphics[width=0.8\linewidth]{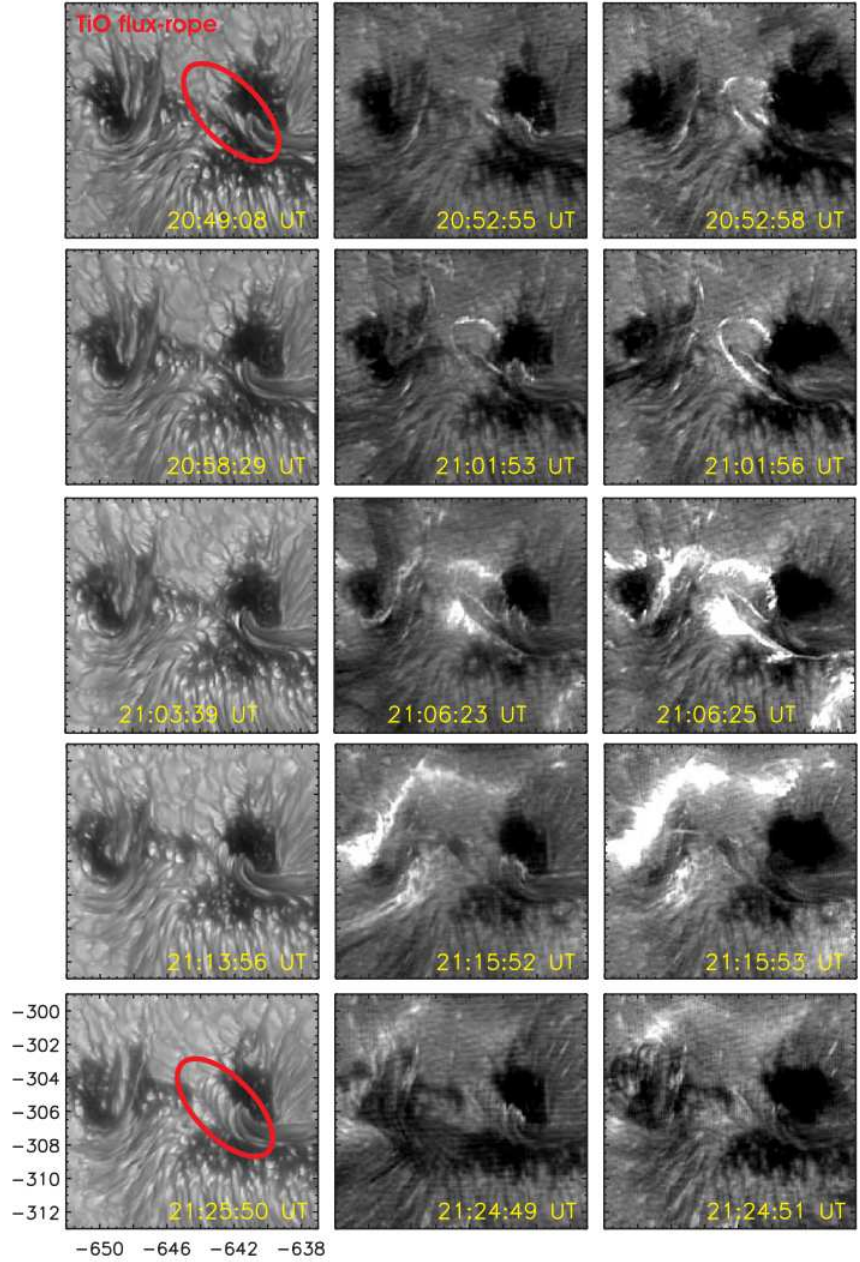}
\caption{Zoomed NST images of the region with highly twisted magnetic field (marked by red arrow in Fig.\,\ref{VIS_large}). TiO images are shown in the left column. The blue wing and red wing H$\rm_{\alpha}$ filtergrams are presented correspondingly in the central and right columns. The twisted TiO magnetic structure called TiO flux-rope is marked by red ellipse in the top-left and bottom-left panels.}
\label{VIS_TIO}
\end{figure}

\begin{figure}[h!]
\centering
\includegraphics[width=0.8\linewidth]{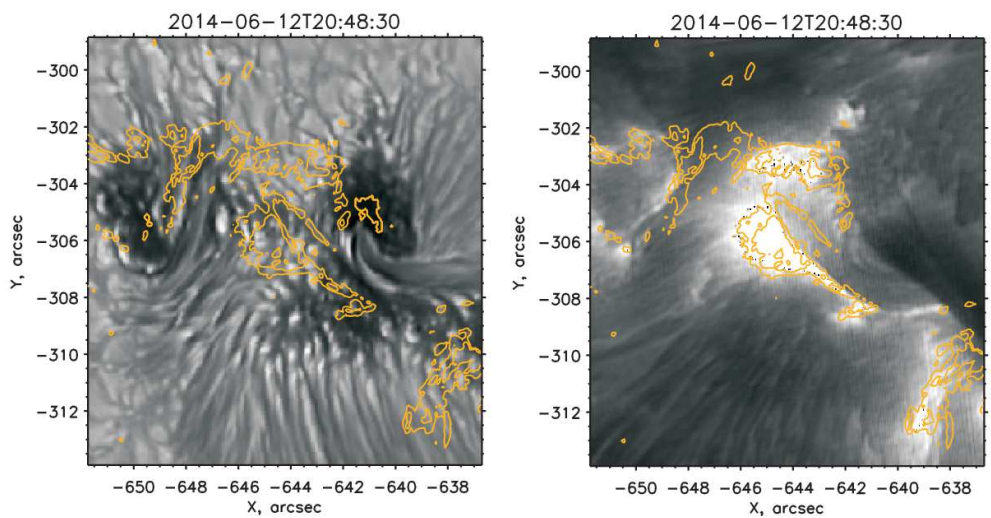}
\caption{The purpose of this figure is to show the correspondence between the H$\rm_{\alpha}$ emission sources and the photospheric structures by overlaying the TiO and H$\rm_{\alpha}$ images. Orange contours correspond to the emission source seen in the H$\rm_{\alpha}$ line red wing filtergrams. Background layers are the corresponding TiO image (left panel), and H$\rm_{\alpha}$ line core filtergrams (right panel).}
\label{VIS_compare}
\end{figure}

More detailed zoomed images of this region are displayed in Fig.\,\ref{VIS_TIO}. The TiO images show the temporal dynamics of an elongated structure (that looks like a small twisted flux rope, or a sheared arcade, and indicated by red ellipse), separating two sunspots of opposite magnetic polarity ($\delta$-type configuration). The visible width of this structure has a tendency to grow with time. Two other columns in Fig.\,\ref{VIS_TIO} correspond to the red and blue wings of $\rm H_{\alpha}$ line. At beginning of the flare before the impulsive phase we observe tiny $\rm H_{\alpha}$ ribbons and brightenings in the vicinity of the TiO arcade. At 21:06:25 UT the most intensive $\rm H_{\alpha}$ ribbons surround the TiO arcade (which may correspond to a magnetic flux rope). A very intriguing peculiarity is that we observe a three ribbon structure: two relatively wide ribbons (with width $\sim 1^{\prime\prime}$) and a tiny thread (width $\sim 0.3^{\prime\prime}$) located between them. The tiny ribbon directly corresponds to the TiO flux-rope which is located inside the twisted magnetic structure, seen in the $\rm H_{\alpha}$ line core. These observations are summarized in Fig.\,\ref{VIS_compare}, where $\rm H_{\alpha}$ (line wing) emission sources (shown by yellow contour lines) are compared with the TiO and $\rm H_{\alpha}$ (line core) images.

\begin{figure}[h!]
\centering
\includegraphics[width=1.0\linewidth]{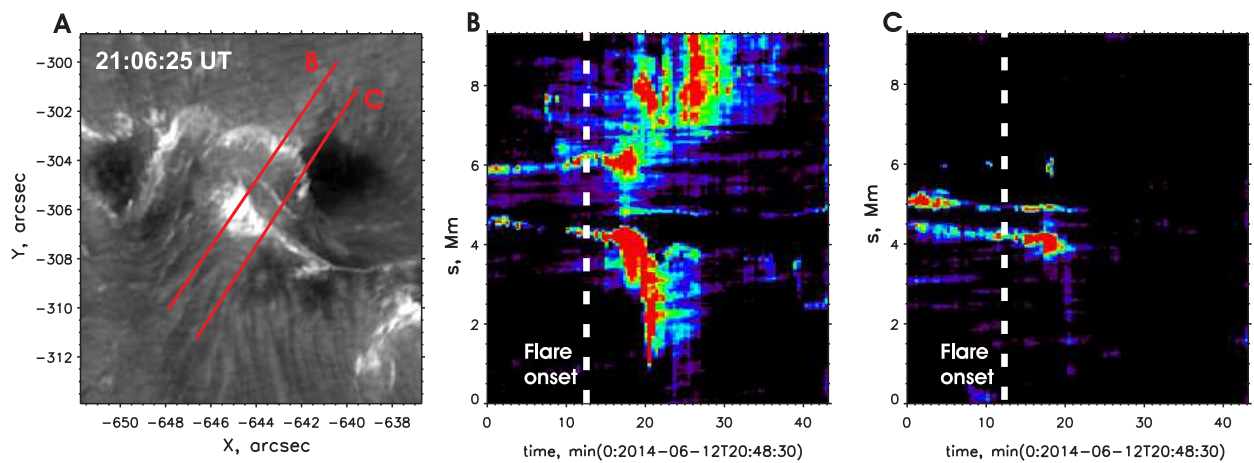}
\caption{Temporal dynamics of the H$\rm_{\alpha}$ (red wing) emission sources. Panels B and C are two variants of the time-space plot with different color contrasts, made along the slices marked in panel A by red line (background in panel A is the H$\rm_{\alpha}$ red wing image).}
\label{Ha_rib_evo}
\end{figure}

A time-space diagram made along two slices intersecting all three $H_{\alpha}$ ribbons is shown in Fig.\,\ref{Ha_rib_evo}. The thin ribbon is quasi-stationary while the other two ribbons experience explosive motion at 21:04~UT in the opposite directions from each other. The approximate speed of this expansion is about 10 km/s for the lower ribbon and $\approx 5$~km/s for the upper ribbon.

\subsection{DYNAMICS OF THE PLASMA FLOWS IN THE FLARE REGION}

In this section we present analysis of the Doppler line-of-sight velocity in the flare region, measured by HMI. It is worth noting that the detected flows do not correspond to the upflows and downflows directly as the solar flare is located $\sim 640^{\prime\prime}$ from the disk center. Thus, estimated velocities are composed from the horizontal (along solar surface) and vertical components. However, further we will distinguish positive and negative velocities as downflows and upflows.

\begin{figure}[h!]
\centering
\includegraphics[width=0.9\linewidth]{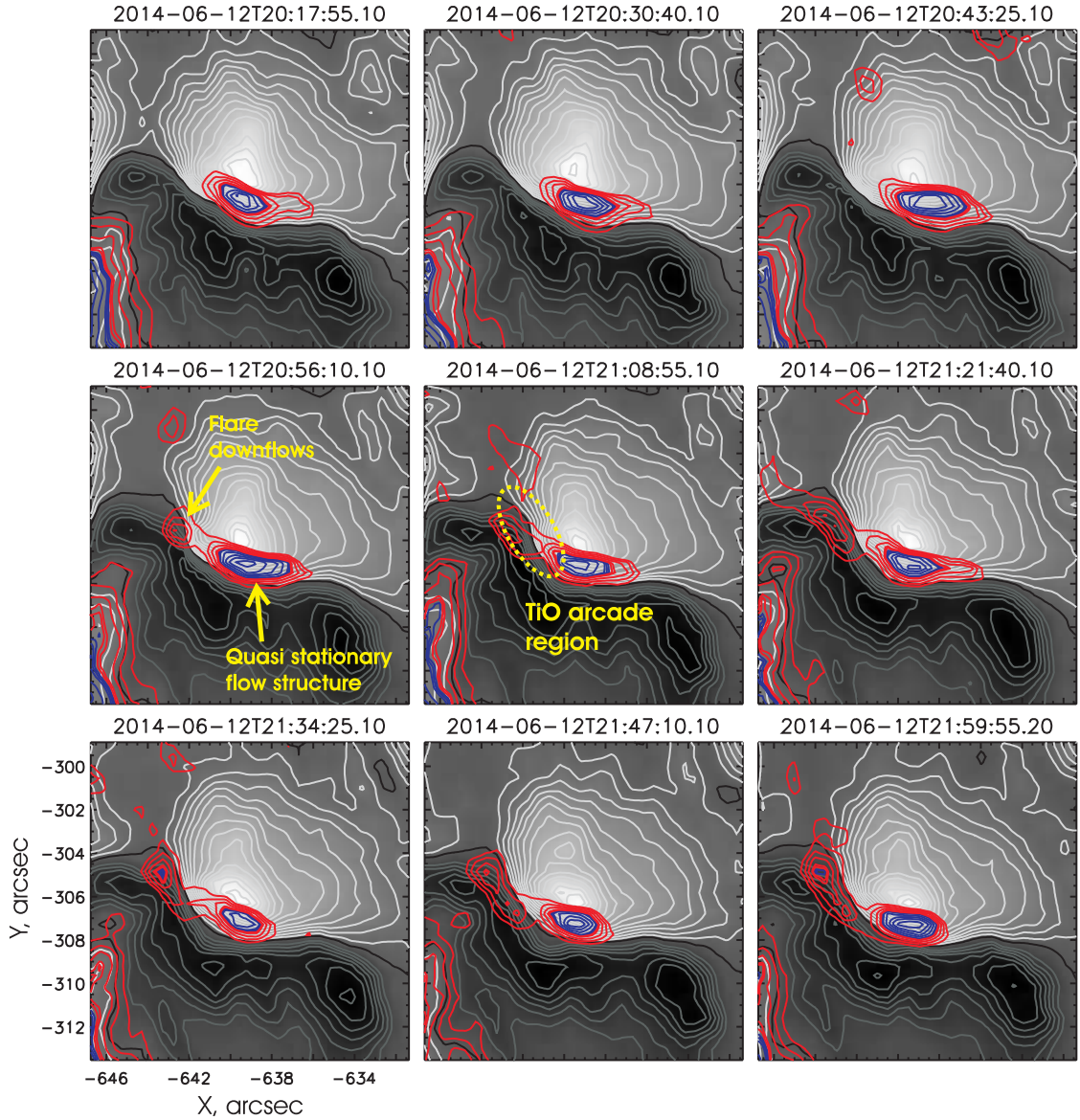}
\caption{Background image is the LOS HMI/SDO magnetogram. Contours mark the levels of Doppler LOS speeds measured by HMI. Upflows with negative speeds are shown by blue contours when downflows with positive speeds are red contours.}
\label{HMI_v}
\end{figure}

The HMI Dopplergrams are compared with the HMI LOS magnetograms in Fig.\,\ref{HMI_v}. Before the flare onset (top rows) we observe a system of upflows surrounded by downflows near the PIL. This flow system was observed during the whole flare process without significant changes. The flare initiation is associated with appearance of new downflows near other region of the PIL (marked by an arrow), which corresponds to the TiO arcade. The average downflows speed is about 1 km/s.

\begin{figure}[h!]
\centering
\includegraphics[width=1.0\linewidth]{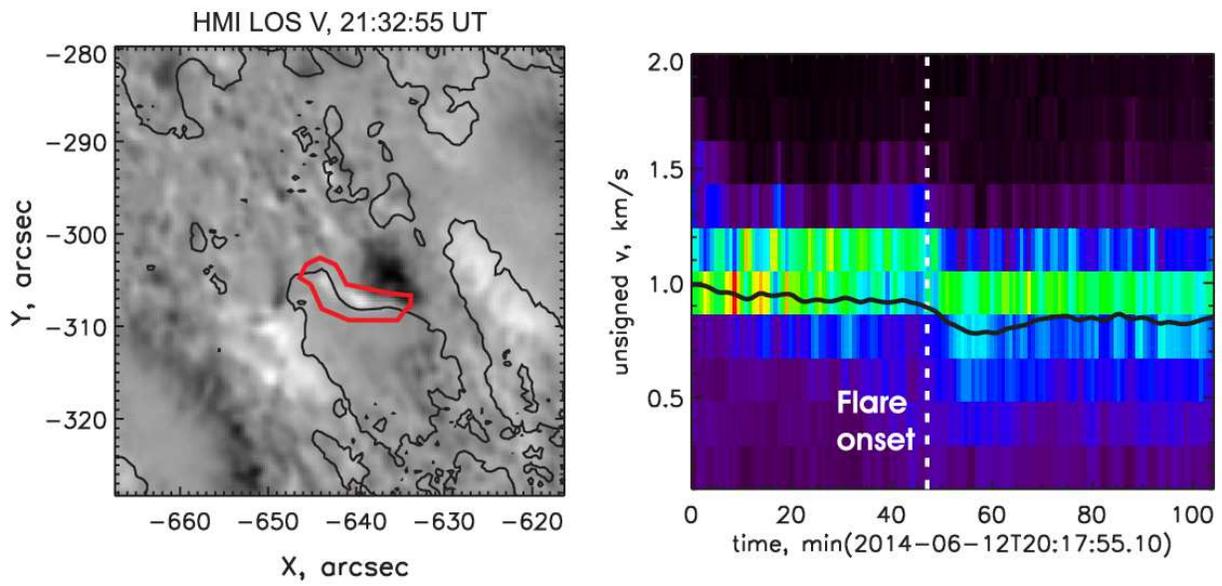}
\caption{The right image presents temporal dynamics of Doppler speed histogram calculated inside the region marked by red contour within the left panel. Background image in the left panel is LOS Dopplergram. Black line marks the LOS PIL.}
\label{hist_v}
\end{figure}

In Fig.\,\ref{hist_v} we demonstrate the evolution of the flow speed distribution in the flare region. By red contour in the image we show an area where distribution of Doppler speeds was calculated. This place corresponds to the TiO arcade and the region with enhanced Doppler flows. One can notice that after the flare onset (marked by vertical line in right panel) the Doppler speed distribution changes. Thick black curve is the Doppler speed averaged over all pixels inside the red contour in left panel. This line also indicates changing flows in the region of our interest which is associated with the flare process.

\subsection{ELECTRIC CURRENTS IN THE FLARE REGION}

To calculate the vertical currents $j_z$ (component perpendicular to the solar surface) we use the HMI vector magnetograms, and the Ampere's law \citep[e.g.][]{Guo2013, Musset2015}:
\vskip-.6cm
\begin{eqnarray}
j_z=\frac{c}{4\pi}(\nabla\times\vec{B})_z = \frac{c}{4\pi}\left(\frac{\partial B_x}{\partial y} - \frac{\partial B_y}{\partial x}\right)
\label{eq_jz}
\end{eqnarray}
where $B_x$ and $B_y$ correspond to the East-West and South-North components of the magnetic field.

\begin{figure}[h!]
\centering
\includegraphics[width=1.00\linewidth]{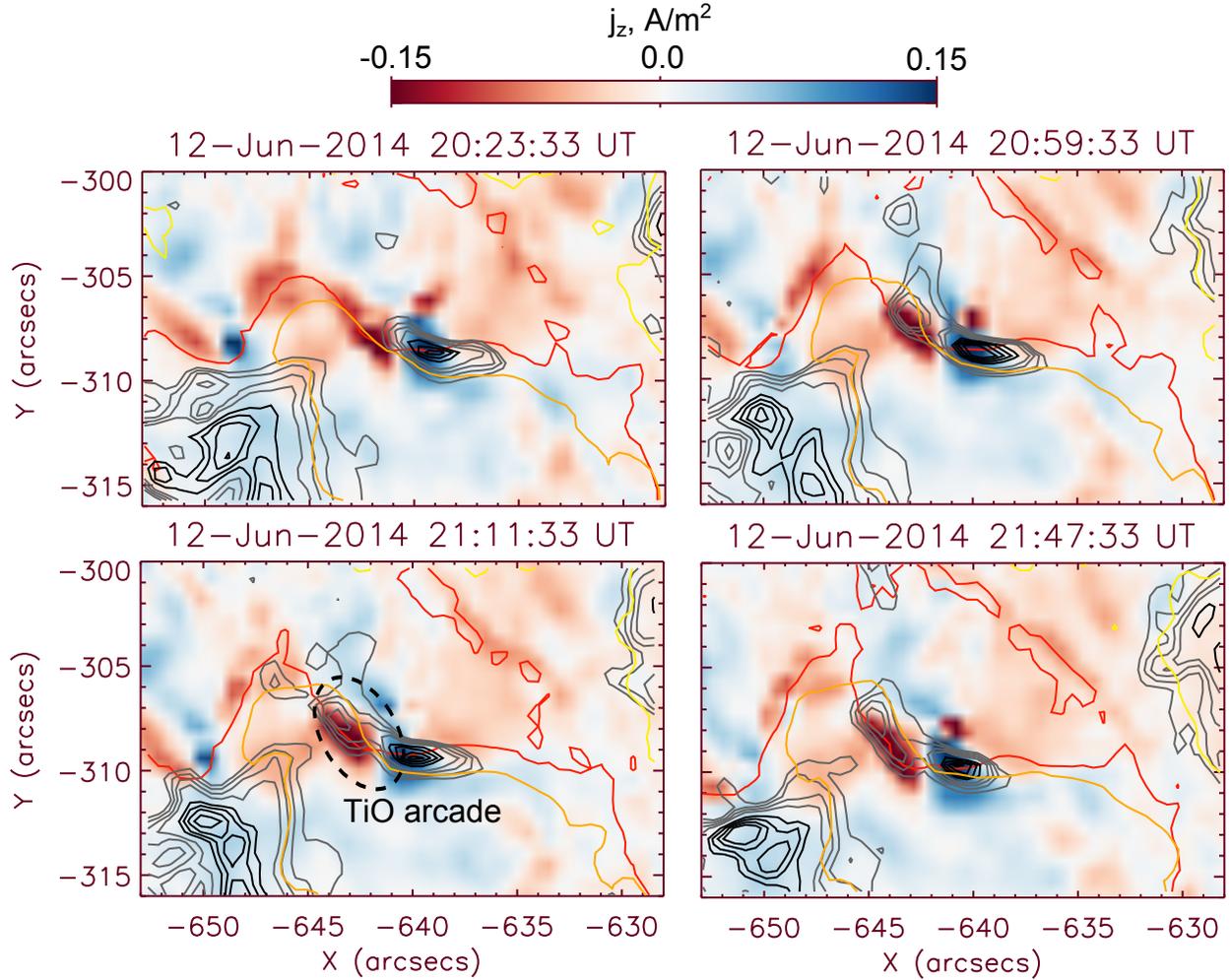}
\caption{Sequence of 4 red-blue images showing the evolution of the vertical component of electric current density. Black and grey contours show upflows (-800, -600, -400 and -200 m/s) and downflows (200, 400, 600, 800 km/s) according to the HMI Doppler measurements. Orange contours show the LOS PIL, while the red lines mark the reconstructed PIL.}
\label{jz}
\end{figure}

Since the flare was located far from the disk center the observed LOS PIL may deviate from the exact position of the PIL of the $B_z$-component (vertical to the solar surface) due to the projection effect. Using the HMI vector magnetograms we recalculated all $\vec{B}$ components from the local helioprojective Cartesian system to the Heliocentric Spherical coordinate system. The recalculated values are used for the $j_z$ calculation (Eq.\,\ref{eq_jz}). Figure~\ref{jz} shows comparison of the LOS PIL (red line) with exact PIL (orange line) in the helioprojective Cartesian coordinate system. Electric currents maps are also shown in Fig.\,\ref{jz}.

\begin{figure}[h!]
\centering
\includegraphics[width=1.0\linewidth]{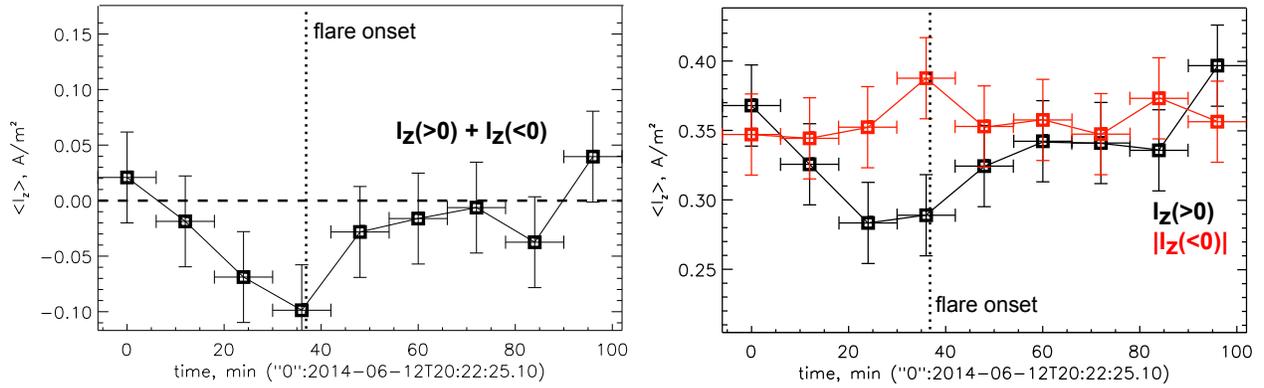}
\caption{Temporal dynamics of total (trough the whole flare region) vertical electric current $I_z$. The left panel show the sum of negative and positive currents, while the right panel presents dynamics of positive (black) and negative (red) $I_z$. Flare onset is marked by vertical dotted line.}
\label{jz_evo}
\end{figure}

There are several places along the PIL where electric currents are intensified. In the region of the TiO flux-rope electric currents are the most intensive. The configuration of electric currents does not experience significant changes which can be visually detected. Temporal dynamics of the averaged $j_z$ is shown in Fig.\,\ref{jz_evo}. The error bars are calculated as standard deviation of the $j_z$ distribution (which is fitted by a Gaussian) outside the region of intensified electric currents near the PIL. One can note that the average current changes during the flare. Absolute value of negative $I_z$ increases while positive $I_z$ decreases.

The reconstructed PIL does not completely correspond to the LOS PIL. Near the TiO flux-rope and downflows region before the flare onset both PILs coincide, but after flare initiation they are separated. This means a change of magnetic field orientation which likely corresponds to the observed reduction of magnetic shear of the TiO arcade, as discussed in the next section.

\subsection{TOPOLOGY OF THE MAGNETIC FIELD}

\begin{figure}[h!]
\centering
\includegraphics[width=1.0\linewidth]{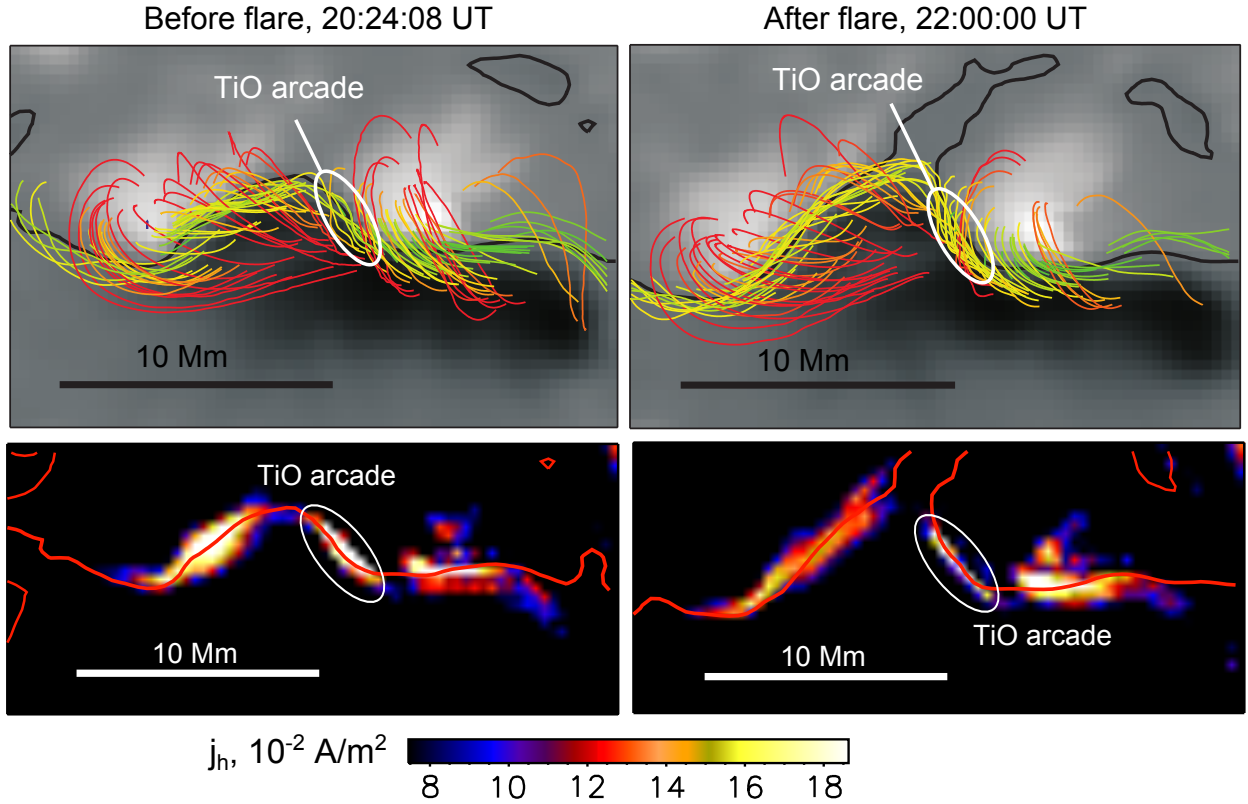}
\caption{The twisted magnetic structure (result of the NLFFF extrapolation) elongated along the PIL (top panels). Different colors mark different heights of the field line (the height increases from green to red color).The left and right panels correspond to the preflare and postflare times. Bottom panels present maps of horizontal component $j_h$ of electric current density, calculated from the NLFFF extrapolations. Black line in top panels and red lines in bottom panels mark the PIL.}
\label{topology}
\end{figure}

The global magnetic field topology of the flare region is shown in Fig.\,\ref{topology} (top panels). The twisted magnetic field structure follows the polarity inversion line of the magnetic field. This twisted structure is surrounded by sheared magnetic field lines forming a closed configuration. The magnetic field topology does not change significantly during the flare energy release, which is shown by comparison between the preflare and postflare NLFFF extrapolations (Fig.\,\ref{topology}).

Bottom panels of Fig.\,\ref{topology} show the absolute value of the horizontal component of electric current density, calculated from the extrapolated magnetic field. We see that the strongest horizontal currents are concentrated in the vicinity of the PIL. One can notice that the TiO arcade corresponds to the place of strong $j_h$. The preflare value of $j_h$ is weaker than in the postflare state.

\begin{figure}[h!]
\centering
\includegraphics[width=1.0\linewidth]{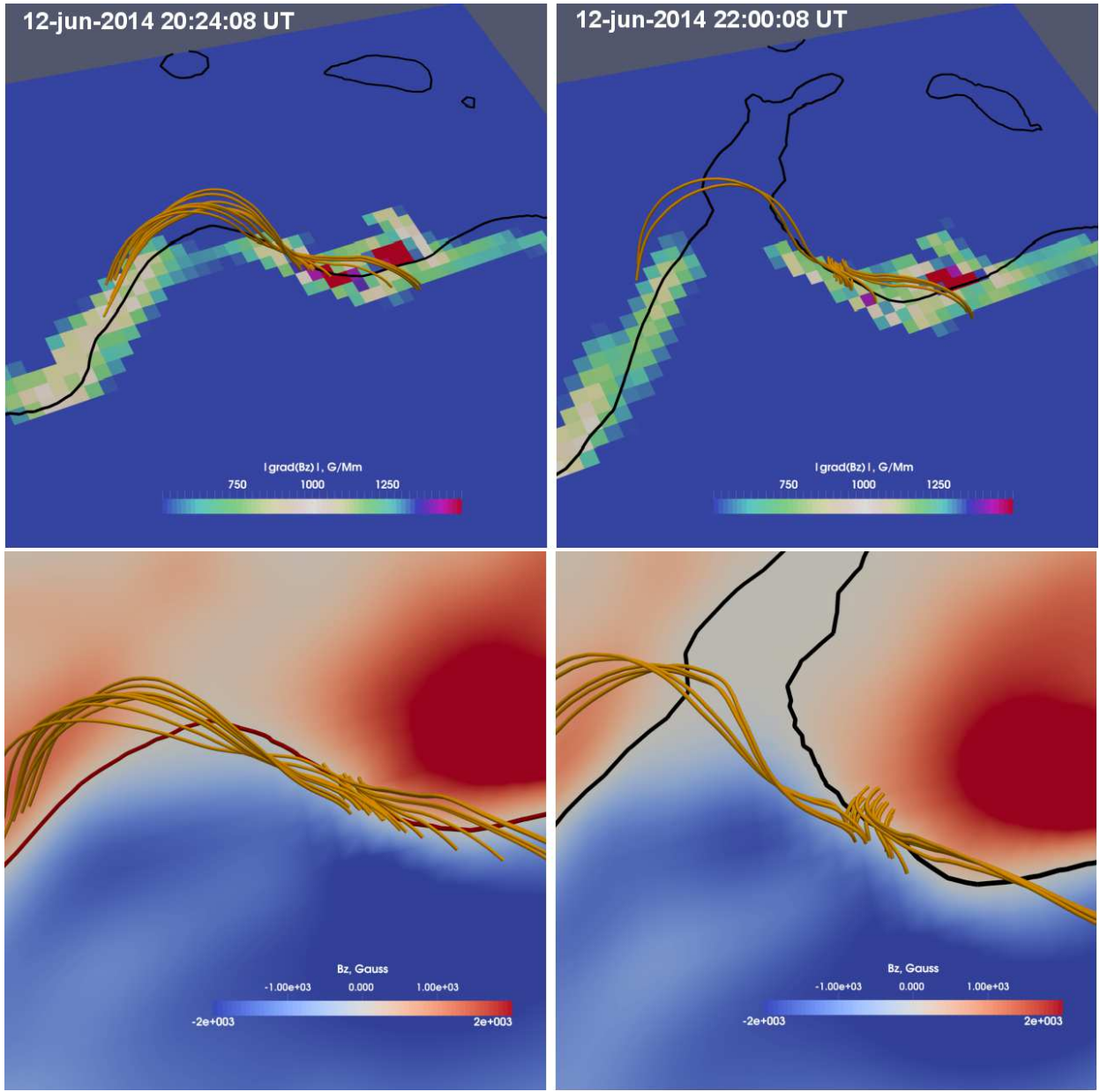}
\caption{Results of the magnetic field extrapolation in the region of TiO flux-rope using NLFFF code. The upper panel is a stereoscopic view with background image corresponding to $\nabla_hB_z$ map. The bottom panels show zoomed image with view from the top, where the background layer is the photospheric magnetogram.}
\label{nlfff2}
\end{figure}

The place of our particular interest is the region, where the TiO arcade and preflare activity were observed (this place is marked by ellipse in Fig.\,\ref{topology}). In Fig.\,\ref{nlfff2} we present the magnetic field extrapolation results for this region. Before the flare, we see two interacting sheared twisted flux-tubes near the magnetic field polarity inversion line. After the flare between these flux-tubes a small arcade is formed. The results of this modelling are consistent with the TiO and $\rm H_{\alpha}$ images (Fig.\,\ref{VIS_TIO}). Before the flare the TiO arcade is more sheared, and after the flare it is more wide and the shear seems to be reduced. In Fig.\,\ref{nlfff2} we also show the value of magnetic field gradient $\nabla_hB_z=\sqrt{(\partial B_z/\partial x)^2+(\partial B_z/\partial y)^2}$ (this value coincides with $j_h$ map in bottom panels of Fig.\,\ref{topology}). The strongest value of $\nabla_hB_z$ ($\sim 1-2$~kG/Mm) is found along the PIL, and concentrated near the region where the TiO flux-rope with downflows was observed. Thus, the strongest shear is along the PIL and TiO arcade.

\begin{figure}[h!]
\centering
\includegraphics[width=0.5\linewidth]{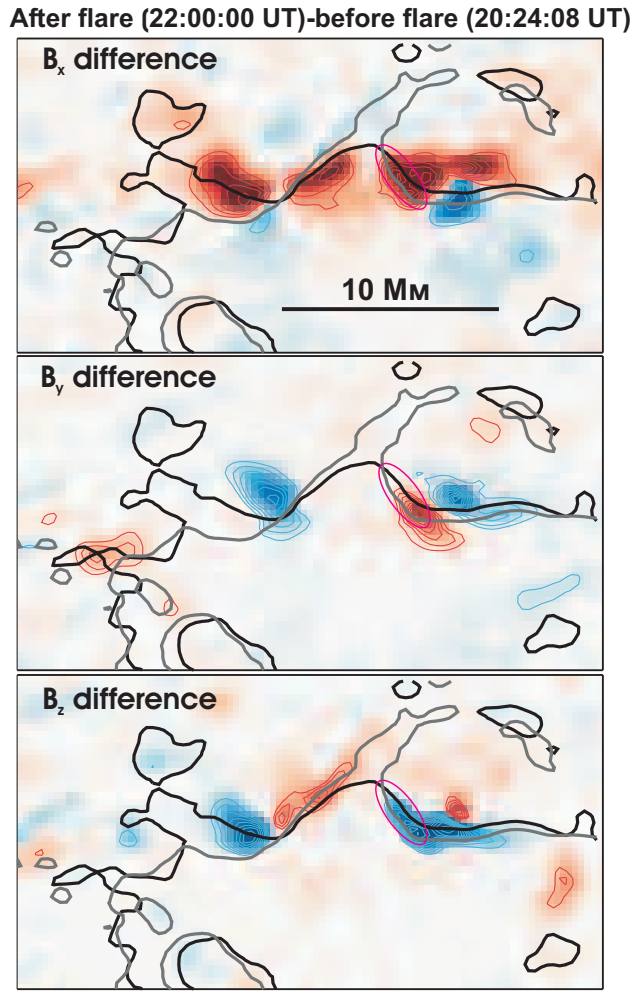}
\caption{Differences between maps of magnetic field components corresponding to preflare (20:24:08 UT) and postflare (22:00:00 UT) times (posflare minus preflare). $B_{x}$, $B_{y}$ and $B_{z}$ difference maps are shown in the top, middle and bottom panels. Black and grey lines mark the preflare and postflare PIL.}
\label{B_restruct}
\end{figure}

To show the dynamics of the magnetic fields in the lower solar atmosphere we calculated differences between the magnetic field components ($B_x$, $B_y$ and $B_z$) before and after the flare (Fig.\,\ref{B_restruct}). One can see that the magnetic field experience changes in the vicinity of the PIL. These changes are especially noticeable in the region of the TiO arcade (indicated by magenta ellipse). The magnetic field dynamics can be associated with magnetic reconnection or motion of the magnetic structures. The motion of the magnetic field tubes produces electric field $\vec{v}\times\vec{B}/c$ which can be an important driver of the flare energy release. Anyway, the significant changes of the magnetic fields observed in the vicinity of the PIL are associated with the flare energy release.

\begin{figure}[h!]
\centering
\includegraphics[width=0.6\linewidth]{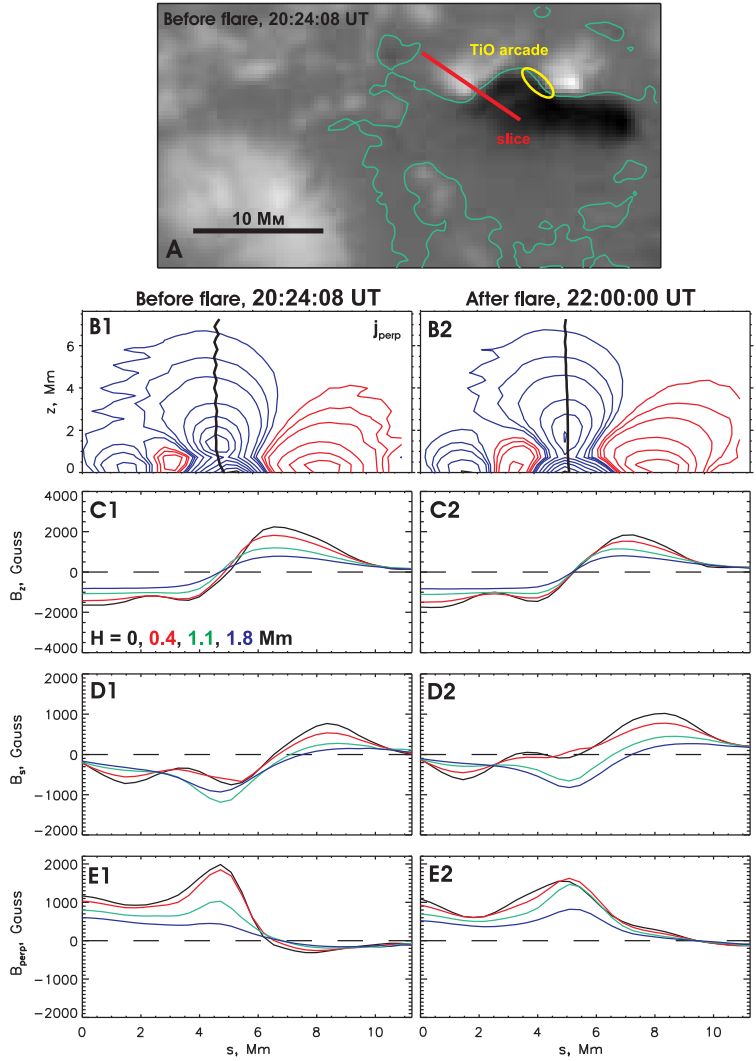}
\caption{Distribution of different physical quantities derived from the NLFFF extrapolations along the slice marked by red line presented in panel A. The background is $B_z$ map and black line is the PIL. Left and right columns correspond to preflare and postflare times. Panels B show $\vec{j}$ component perpendicular to the slice (black line is the PIL in the plane of slice). Blue contours correspond to positive (directed to the reader) $j_{perp}$ with levels 6, 11, 19, 37, 56, 75, 93, 112, 130 and 150 mA/m$^2$. Red contours correspond to negative $j_{perp}$ with levels 6, 11, 19, 37, 56 and 75 mA/m$^2$. Panels C, D, and E show $B_z$, $B_s$ and $B_{perp}$ along the slice at different heights.}
\label{slice1}
\end{figure}

\begin{figure}[h!]
\centering
\includegraphics[width=0.6\linewidth]{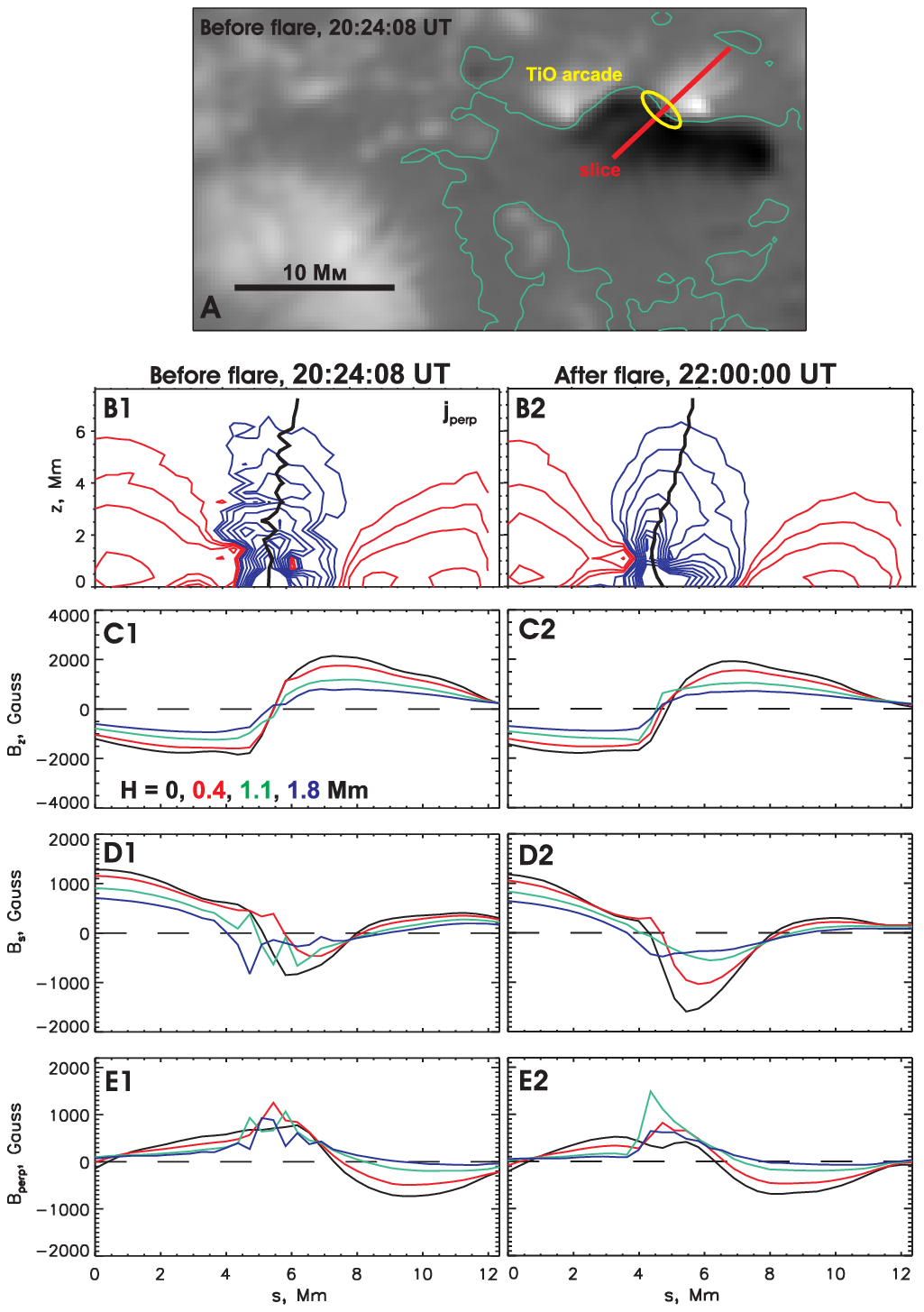}
\caption{The same physical quantities as in the Fig.\,\ref{slice1} but made for another slice (red line in the panel~A). Blue contours correspond to positive $j_{perp}$ with levels 6, 11, 19, 37, 56, 75, 93, 112, 130 and 150 mA/m$^2$. Red contours correspond to positive $j_{perp}$ with levels 6, 11, 19, 37, 56 and 75 mA/m$^2$.}
\label{slice2}
\end{figure}

\begin{figure}[h!]
\centering
\includegraphics[width=0.6\linewidth]{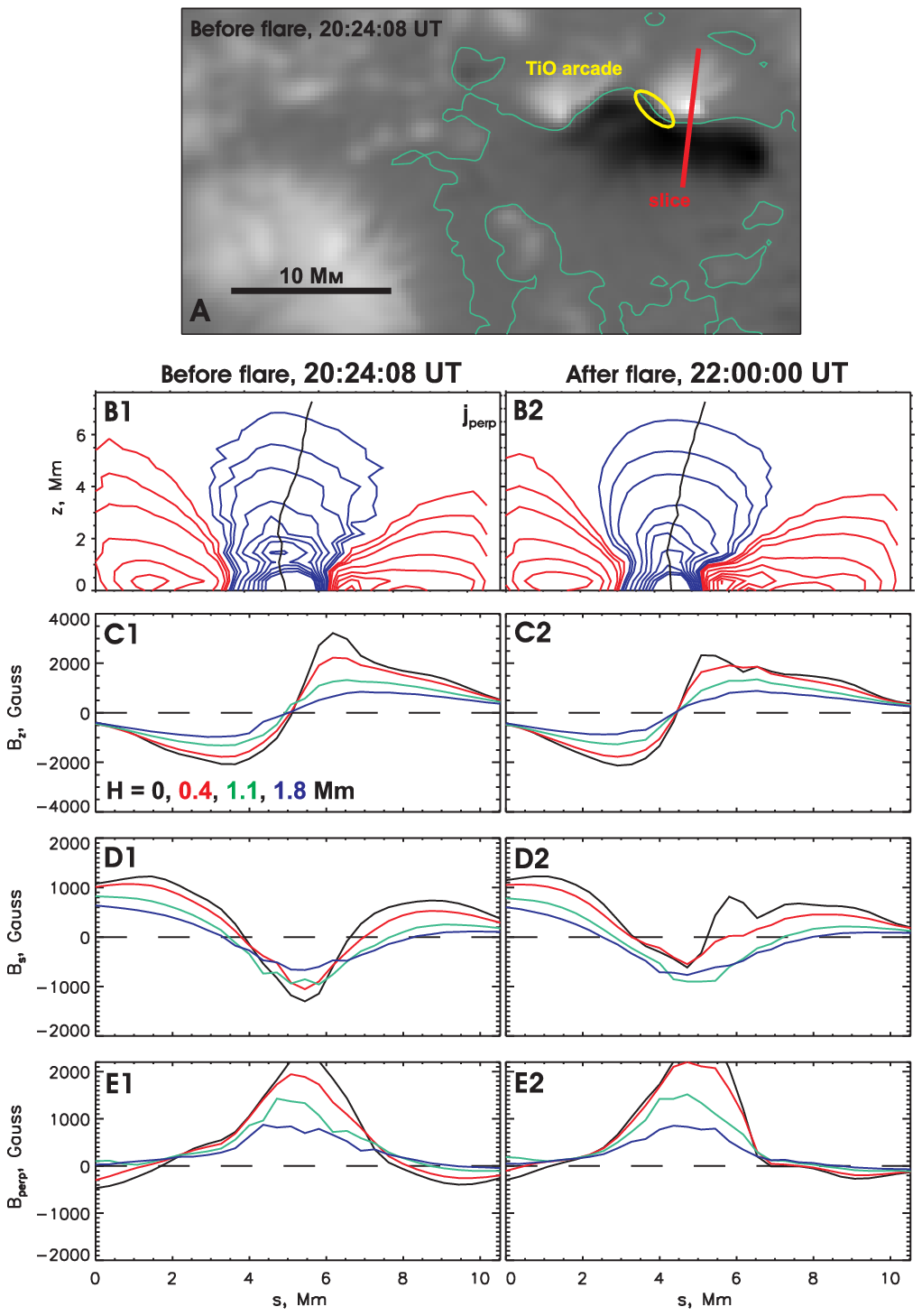}
\caption{The same physical quantities as in the Fig.\,\ref{slice1} but made for another slice (red line in the panel~A). Blue contours correspond to positive $j_{perp}$ with levels 6, 11, 19, 37, 56, 75, 93, 112, 130 and 150 mA/m$^2$. Red contours correspond to positive $j_{perp}$ with levels 6, 11, 19, 37, 56, 75, 93, 112 and 130 mA/m$^2$.}
\label{slice3}
\end{figure}

In Figures~\ref{slice1}-\ref{slice3} we demonstrate the distribution of different physical quantities derived from the NLFFF magnetic extrapolations along three slices intersecting the PIL in various points (Fig.\,\ref{slice2} show slice intersecting the TiO arcade). We show these distributions for preflare (left panels) and postflare (right panels) times. There are no any significant changes in the distribution of the electric current component $j_{perp}$ perpendicular to the slice.

Magnetic field components $B_z$ (panel E), $B_s$ (along slice, panel F) and $B_{perp}$ (perpendicular to the slice, panel G) at various heights also does not show significant changes during the flare energy release. We only found increasing $B_s$ component in the case of slice presented in Fig.\,\ref{slice2} and decreasing $B_{s}$ in the other two cases (Figures~\ref{slice1} and \ref{slice3}).

We also show the sign inversion line of $B_z$ component in the plane of the slice by black line (panels B and C). This line experiences some changes only in the region of the TiO arcade (Fig.\,\ref{slice2}). In the case of other slices (Figures~\ref{slice1} and \ref{slice3}) the line does not change during the flare energy release.

\begin{figure}[h!]
\centering
\includegraphics[width=0.6\linewidth]{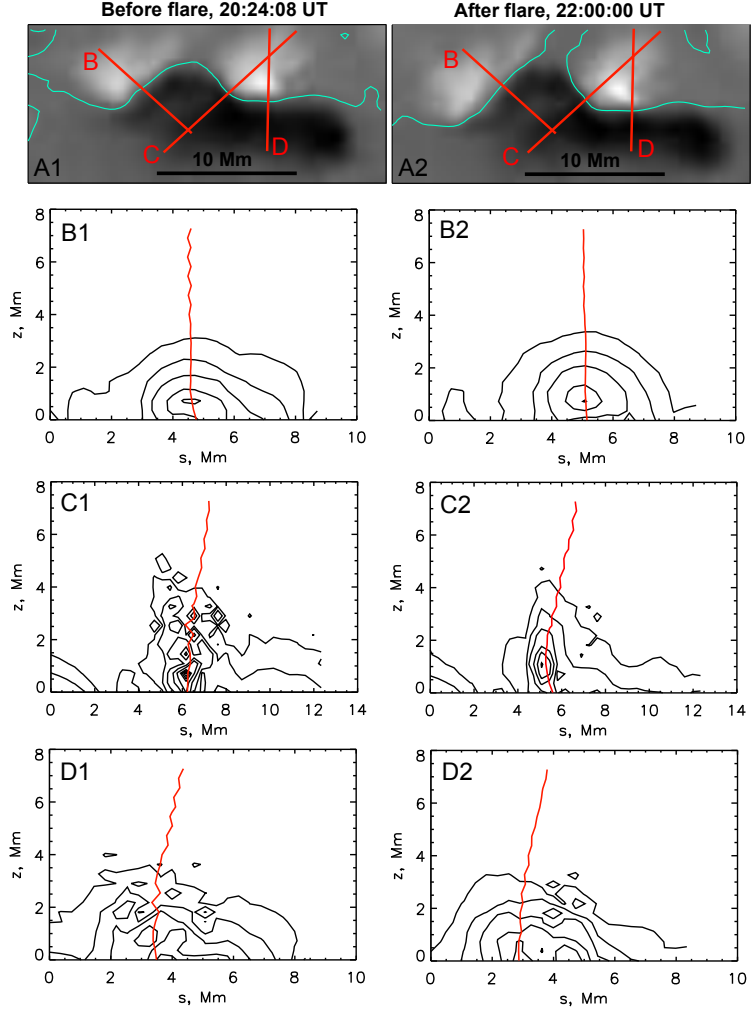}
\caption{Figure shows changes in the distribution of absolute value of $\vec{j}$ along the slices B, C, D shown in two top panels with $B_z$ map at preflare (A1) and postflare (A2) times. Panels B, C and D correspond to slices B, C and D. Contours mark electric current density levels with values: 15, 37, 75, 150, 224, 298, 336 and 373 mA/m$^2$. Red line is the PIL in the plane of slice.}
\label{jslices}
\end{figure}

In Figure~\ref{jslices} we show the distribution of $\vec{j}$ absolute value. One can note that in the region of the TiO arcade we have decreasing value of electric currents and the height of the maximal electric current is decreased. While other slices do not show significant changes.

\section{DISCUSSION AND CONCLUSIONS}

In this section we will try to draw a picture of the physical processes in the flare region. First of all, we summarize the main observational results. All preflare activity in the low solar atmosphere was located near the PIL in a compact region, according to the NST data. The flare observed in the NST images was accompanied by changing magnetic fields near the PIL which is clearly seen in the HMI vector magnetograms. The TiO and H$\rm_{\alpha}$~images reveal formation of a small ($\sim 3$Mm) arcade-like magnetic structure in the photospheric and chromospheric layers of the solar atmosphere, which shows an evolving shearing structure. Magnetic field of this structure is approximately $1000$ Gauss. The H$\rm_{\alpha}$ emission sources near this place were observed as a three ribbon structure with the thinnest ribbon located between the thicker ones. This thin ribbon was stable while the others moved away from each other during the flare with the speed $\sim 5-10$ km/s. Moreover the thin ribbon was observed only in the wings of H$\rm_{\alpha}$ line, that evidences in favour of a deep location of this structure.

In the same region of the small magnetic arcade we found intensification of Doppler downflows across the PIL, which is associated with the flare onset. The strongest intensification of electric currents estimated from the HMI vector magnetograms also corresponds to this place. Downflows could be connected with moving plasma attached to the reconnected magnetic field lines. The plasma flows across magnetic field lines produce electric field $\vec{v}\times\vec{B}/c$ which could be a driver of the flare energy release. The observed flows near the PIL are connected with the flare energy release as their distribution and average speed experiences changes during the flare time.

\begin{figure}[h!]
\centering
\includegraphics[width=1.0\linewidth]{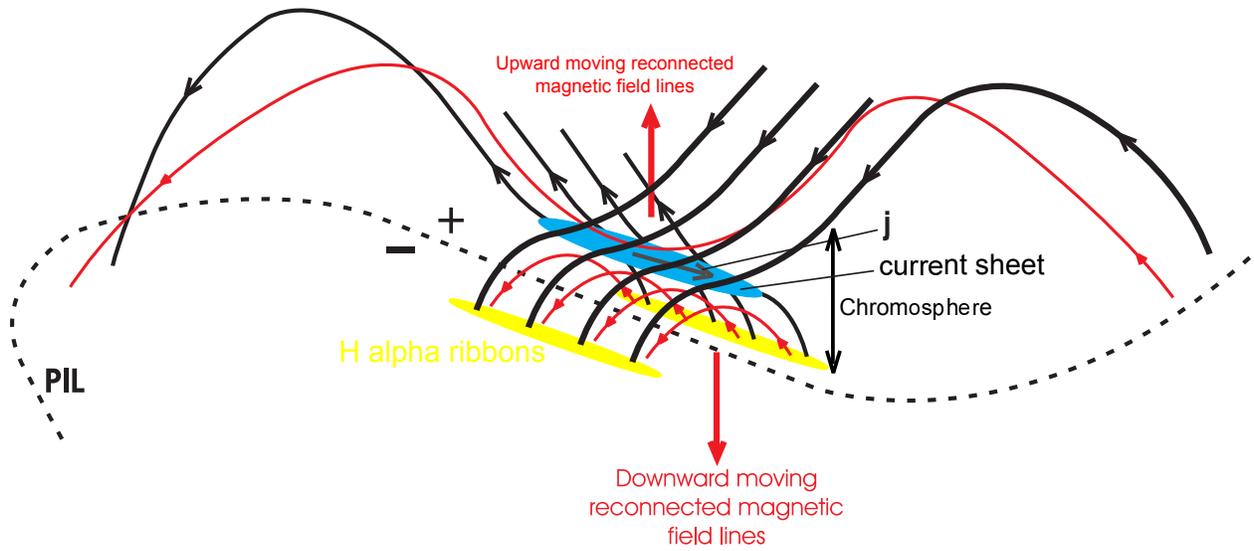}
\caption{The figure presents a cartoon schematically showing the magnetic reconnection process in the chromosphere which probably takes place in the studied flare.}
\label{scheme}
\end{figure}

The NLFFF modelling reveals interaction of oppositely directed magnetic flux-tubes in the PIL. The strong electric current was concentrated near the PIL and its strongest value is achieved in the region of the TiO arcade. Before the flare onset the interaction of the magnetic fluxes extended from the photospheric up to the chromospheric layers. These two interacting magnetic flux tubes are observed as the sheared TiO arcade. After the flare the arcade shear is reduced in accordance with the NST observation. A possible scheme of the magnetic reconnection and magnetic field topology in the PIL is illustrated in Fig.\,\ref{scheme}. The reconnection is developed in the volume with strong magnetic field in the chromosphere codirectional with the electric current. Footpoints of the formed arcade may correspond to the observed H$\rm_{\alpha}$ ribbons. How can we explain the appearance of the third thin ribbon located between the thick ones in the low solar atmosphere, elongated along the PIL, and cospatial with the TiO arcade? This H$\rm_{\alpha}$ emission could originate from a thin channel (possibly a part of the current sheet) where electric current dissipates.

How can we explain formation of the large scale emission ribbons? Can the energy be transferred from site in the PIL in the lower atmosphere to the flare ribbons? Perhaps, it is possible to organise energy transfer by accelerated particles spreading along the magnetic field lines, and filling the large-scale magnetic field structure. Heat conduction flows can be also responsible for energy transfer. However, to answer these questions theoretical modeling is necessary.




Presented observational results evidence in favour of location of the primary energy release site in the dense chromosphere where plasma is partially ionized in the region of strong electric currents concentrated near the polarity inversion line. Magnetic reconnection may be triggered by two interacting magnetic flux tubes with forming current sheet elongated along the PIL. The reconnection process develops in the presence of strong magnetic field $\sim 1000$ Gauss and, probably, large radiative losses due to the high density. Observed small magnetic arcade can be a result of the chromospheric magnetic reconnection.

The studied solar flare is not eruptive and, thus, energy of CME should not be considered in total flare energetics. To calculate total flare energetics one can integrate flare emissions. Total integrated soft X-ray radiation losses are estimated as $\sim 10^{30}$ ergs according to the GOES data. This energy was estimated by time integration of $L_{rad}=EM\cdot f(T)$, where $f(T)$ is radiative loss function \citep{Rosner1978}. Emission measure $EM$ and plasma temperature $T$ are calculated according to the method described in the work of \cite{Thomas1985}. However, flare radiation losses are concentrated not only in the soft X-ray emission range. Significant part of the flare radiation is also concentrated in the UV and EUV parts of electromagnetic spectrum \citep{Emslie2012,Milligan2014}. According to these works, the total flare bolometric irradiance can be several times greater than the soft X-ray integrated energy losses. In the case of the studied flare, we may suggest that total flare energetics $E_{tot}$ is in the range of $10^{30}-10^{31}$ ergs. This value can be explained by dissipation of the energy stored in the magnetic fields, generated by local electric currents. The total magnetic energy is calculated as integration of $B^2/8\pi$ over the whole volume, for which we made the magnetic field extrapolations. According to the NLFFF modelling the difference between the postflare and preflare total magnetic energies is about $6\times 10^{31}$ ergs which is, at least, a few times larger than energy $E_{tot}$ released during the studied flare. Thus, the released magnetic energy is in accordance with the total flare energetics.

The studied event is a good illustration of a chromospheric non-eruptive flare which does not fit into a standard model. It motivates us to investigate magnetic reconnection in the physical conditions far from those that are assumed in the frame of the standard model. To improve idea of flare chromospheric magnetic reconnection we need new detailed numerical simulations whose results can be compared with the available high-resolution multiwavelength observations of the solar flares.

The main conclusion of the paper is that the flare primary energy release develops within a relatively small volume in the vicinity of the magnetic field polarity inversion line in the presence of strong electric currents, magnetic fields and plasma flows confirming the initial results of \cite{Severnyi1958}. It is likely that all large scale processes of flare energy release observed as UV, EUV and $\rm H_{\alpha}$ ribbons and formation of hot UV and soft X-ray loops are connected with the small scale primary energy release site located near the PIL.

\acknowledgements
The authors acknowledge the BBSO observing and technical team for their contribution and support. This work was partially supported by the Russian Foundation for Basic Research (grants 15-32-21078 and 16-32-00462), and by NASA grant NNX14AB68G.



\clearpage



\clearpage
\end{document}